\RequirePackage{fix-cm}
\documentclass[smallextended]{svjour3}       % onecolumn (second format)
\smartqed  % flush right qed marks, e.g. at end of proof
\usepackage{graphicx}
\journalname{Experimental Astronomy}

\begin{document}
\title{Re-testing the JET-X Flight Module No. 2 at the PANTER facility}
\titlerunning{Re-testing the JET-X Flight Module No. 2 at the PANTER facility}        % if too long for running head

\author{Daniele Spiga, Gianpiero Tagliaferri, Paolo Soffitta, Oberto Citterio, Stefano Basso, Ronaldo Bellazzini, Alessandro Brez, Wolfgang Burkert, Vadim Burwitz, Enrico Costa, Luca de Ruvo,  Ettore Del Monte, Sergio Fabiani, Gisela Hartner, Benedikt Menz, Massimo Minuti, Fabio Muleri, Giovanni Pareschi, Michele Pinchera, Alda Rubini, Carmelo Sgr\`o, Gloria Spandre}

\authorrunning{D. Spiga, G. Tagliaferri, S. Basso, O. Citterio, et al.} % if too long for running head

\institute{D. Spiga\at
              INAF / Brera Astronomical Observatory, Via Bianchi 46, 23807, Merate (Italy) \\
              Tel.: +39-039-5971027\\
              Fax: +39-039-5971001\\
              \email{daniele.spiga@brera.inaf.it}           %  \\
           \and
           G. Tagliaferri, S. Basso, O. Citterio, G. Pareschi\at
           INAF / Brera Astronomical Observatory, Via Bianchi 46, 23807, Merate (Italy)
           \and
           E. Costa, E. Del Monte, S. Fabiani, F. Muleri, P. Soffitta, A. Rubini\at
           INAF/Institute for Space Astrophysics and Planetology, via Fosso del Cavaliere 100, 00133 Roma (Italy)
           \and
           R. Bellazzini, A. Brez, L. de Ruvo, M. Minuti, M. Pinchera, C. Sgr\`o, G. Spandre\at
           INFN Sezione di Pisa, Largo B. Pontecorvo, 3 - 56127 Pisa (Italy)
           \and
           V. Burwitz, W. Burkert, G. Hartner, B. Menz\at
           Max-Planck-Instit\"ut f\"ur extraterrestrische Physik, Gie$\beta$enbachstra$\beta$e 1, 85748 Garching (Germany)
}
\date{Received: 1 September 2013 / Accepted: date}

\maketitle

\begin{abstract}
The Joint European X-ray Telescope (JET-X) was the core instrument of the Russian Spectrum-X-$\gamma$ space observatory. It consisted of two identical soft X-ray (0.3 -- 10 keV) telescopes with focusing optical modules having a measured angular resolution of nearly 15~arcsec. Soon after the payload completion, the mission was cancelled and the two optical flight modules (FM) were brought to the Brera Astronomical Observatory where they had been manufactured. After 16 years of storage, we have utilized the JET-X FM2 to test at the PANTER X-ray facility a prototype of a novel X-ray polarimetric telescope, using a Gas Pixel Detector (GPD) with polarimetric capabilities in the focal plane of the FM2. The GPD was developed by a collaboration between INFN-Pisa and INAF-IAPS. In the first phase of the test campaign, we have re-tested the FM2 at PANTER to have an up-to-date characterization in terms of angular resolution and effective area, while in the second part of the test the GPD has been placed in the focal plane of the FM2. In this paper we report the results of the tests of the sole FM2, using an unpolarized X-ray source, comparing the results with the calibration done in 1996.
%\subclass{85-05}
\end{abstract}

\section{Introduction}
\label{intro}
JET-X (Joint European X-ray Telescope \cite{Wells92}) was the core imaging instrument onboard the Spectrum-X-$\gamma$ space observatory, stemming from an international collaboration between Italy (Brera Astronomical Observatory, CNR institutes in Milano and Palermo, University groups in Milano and Rome), United Kingdom (University of Leicester and Birmingham, the Rutherford Appleton Laboratory and the Mullard Space Science Laboratory), Germany (Max-Planck-Instit\"ut f\"ur extraterrestrische Physik, MPE), and ESA-ESTEC. JET-X consisted of two identical, co-aligned telescopes with sensitivity in the soft X-ray band (0.3 - 10 keV) and an angular resolution requirement of better than 30 arcsec HEW (Half-Energy-Width, the angular diameter enclosing 50\% of the focused X-rays). The optical modules \cite{Citterio94} were developed under contract of ASI (the Italian Space Agency) and manufactured using the technique of mandrel replication by Nickel electroforming \cite{Citterio96}. This methodology enables the production of grazing-incidence X-ray mirrors with low thickness (down to 0.3 mm) maintaining a very good angular resolution (HEW $<$ 15 arcsec), therefore twice better than the initial requirement. Thin mirror walls allow for a reduction in mass of a single mirror, but also allow one to nest many co-axial and co-focal mirrors to increase the module effective area. The electroforming technique, developed at the CNR in Milano and subsequently at the Brera Astronomical Observatory, was also successfully used to manufacture the optical modules of Beppo-SAX \cite{SAX}, XMM-Newton \cite{XMM}, and is also adopted to produce the 7 optical modules of the X-ray telescope eROSITA \cite{Friedrich}, onboard an observatory -- also called Spectrum-X-$\gamma$ -- to be launched in 2014.

The two optical Flight Modules (FM1 and FM2) and the spare module (FM3) comprise 12 grazing incidence mirror "shells" (Tab.~\ref{tab:JETXfacts}) with a Wolter-I axial profile, commonly used in X-ray optics, consisting of a paraboloid and a hyperboloid \cite{VanSpey}, onto which the rays make two consecutive reflections. The mirror shells have been replicated from Wolter-I mandrels, which were precisely figured and polished at the C. Zeiss company in Oberkochen (Germany). The Gold reflective layer was directly deposited onto the mandrels by e-beam evaporation at the Media-Lario company (Bosisio Parini, Italy), where the Nickel shells were subsequently electroformed and assembled into the optical modules. The JET-X FMs were tested in X-rays \cite{Citterio96} at the PANTER facility (MPE Neuried, Germany \cite{Burwitz}), returning effective area performances very close to the theoretical expectations, and HEW values near 15 arcsec at 1.5~keV. Also end-to-end tests were completed at PANTER, thereby demonstrating the compliance of JET-X to the design specification \cite{Wells}.

\begin{figure}
  	\includegraphics[width=0.75\textwidth]{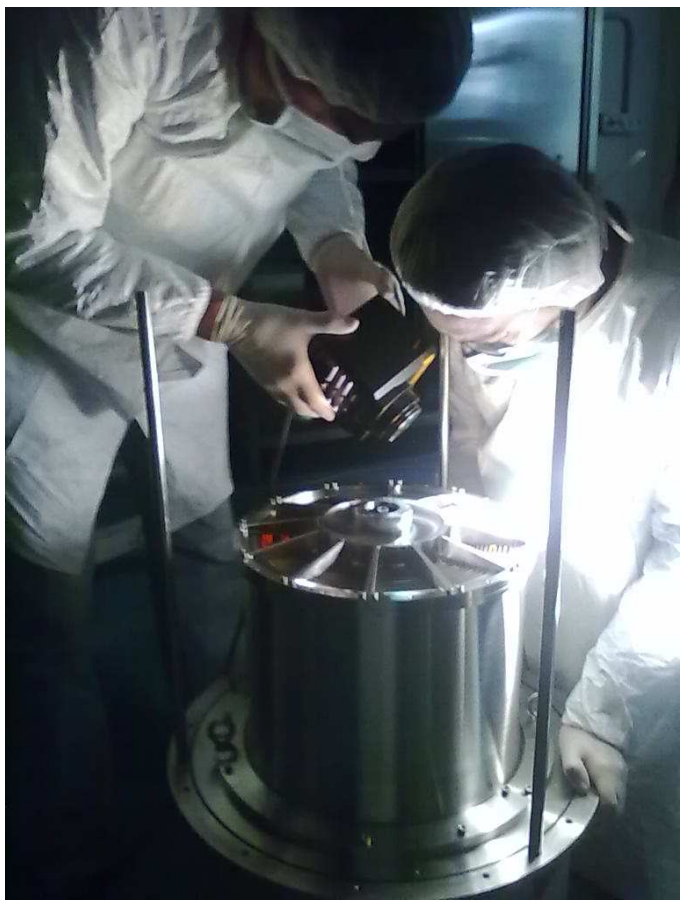}
	\caption{Inspection of the JET-X FM2 before tests at PANTER. After 16 years of storage in Nitrogen, the mirror fixtures on the spider are checked with a high intensity lamp, in a clean room environment.}
	\label{fig:inspection1}  
\end{figure}

The original Spectrum-X-$\gamma$ mission was cancelled in 2002. Consequently, the FM1 and the FM2 did not fly, whilst the FM3 spare module was re-utilized to serve as imaging optic of the X-Ray Telescope (XRT, \cite{Moretti04}, \cite{Tagliaferri}) onboard the SWIFT observatory \cite{Burrows}, successfully launched in 2004 and in operation until today. Meanwhile the JET-X FMs, property of ASI, were dismounted from the payload and brought back in 2008 to the Brera Astronomical Observatory, where they were kept stored in Nitrogen atmosphere.

\begin{figure}
 	 \includegraphics[width=0.75\textwidth]{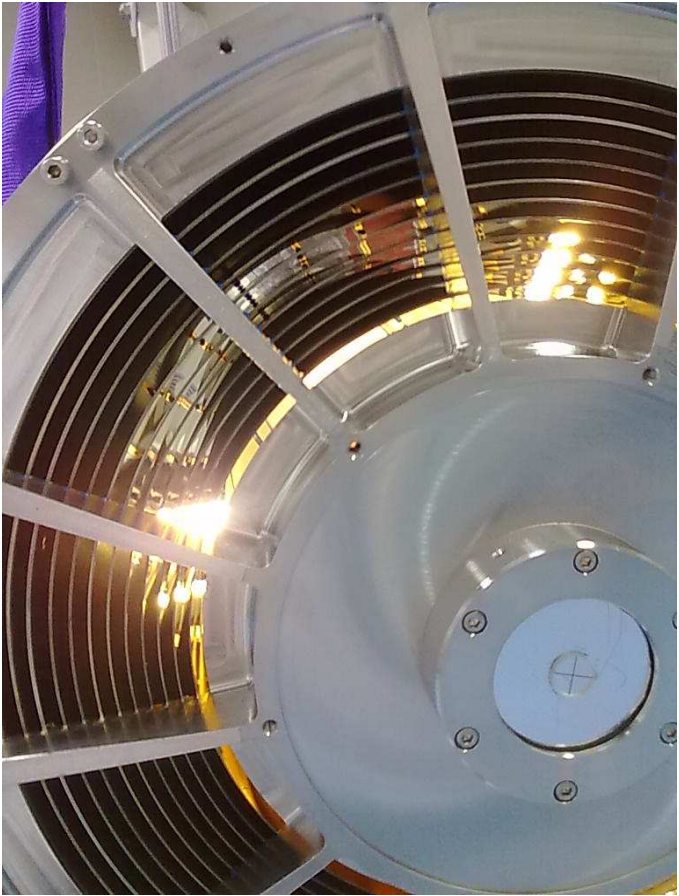}
	\caption{Inspection of the JET-X FM2 before tests at PANTER. The reflecting surface cleanliness is checked with a high intensity light.}
	\label{fig:inspection2}      
\end{figure}

In fall 2012 we have carefully extracted the JET-X FM2 out of its storage case (Fig.~\ref{fig:inspection1} and Fig.~\ref{fig:inspection2}), to perform at PANTER the first experiment of X-ray imaging polarimetry \cite{Fabiani}, hence coupling a high angular resolution focusing optic to a Gas Pixel Detector (GPD) sensitive to the X-ray polarization \cite{Muleri} developed by a collaboration between INFN-Pisa and INAF-IAPS. The scientific interest of this experiment is related to the urgent need of {\it sensitive, well-calibrated, and spatially-resolving} X-ray polarimetry observatories in order to probe the X-ray source models and ascertain the physical processes at work in compact X-ray sources (cyclotron, synchrotron, aspheric accretion, magnetized plasmas, strong and Quantum Gravity effects \cite{Costa}). However, the field of X-ray polarimetry remains so far unexplored, even if polarization-devoted mission concepts (POLARIX \cite{Costa}, and more recently XIPE \cite{Soffitta}) were proposed and studied in the last years, but not selected by the space agencies. Also the NHXM hard X-ray observatory, devoting one of its 4 optical modules to imaging polarimetry \cite{Tagliaferri_NHXM}, has been proposed to ESA but not selected. 

The experiment reported in this work and in \cite{Fabiani} demonstrates for the first time the feasibility of an X-ray imaging polarimeter, testing an X-ray imaging polarimeter prototype obtained by mounting at the PANTER facility the JET-X FM2 and a GPD in its focal plane. In this paper we report the results of the first part of the campaign, addressed to re-testing the angular resolution and the effective area of the JET-X FM2 at some selected monochromatic X-ray energies. In fact, despite the careful storage, we cannot in principle exclude a degradation of the optical performances, e.g., a coating aging, or a stress relaxation in the mirror walls, although very unlikely (the FM3 is in orbit since 2004 and is still operating to date). We describe the experimental setup at PANTER in Sect.~\ref{setup}. In Sect.~\ref{results} we show the experimental results, we compare the results achieved with the previous ones and with some modeling of the effective area and the angular resolution variation with the X-ray energy. In this way, we could infer a power spectrum for the mirror roughness and find evidence for a very thin hydrocarbon contamination layer over the Gold surface.

The results of the second part of the campaign, in which the GPD is used to image the focal spot of the JET-X FM2 and measure the angular resolution of the resulting system (using the same source setup) is described in detail in \cite{Fabiani}. Tests with polarized radiation will be performed at a subsequent time to measure the polarization sensitivity of the FM2+GPD polarimetric system.

\begin{table}
\caption{Geometric characteristics of the JET-X FM2.}
\label{tab:JETXfacts} 
	\begin{tabular}{ll}
		\hline\noalign{\smallskip}
		Nominal focal length & 3500 mm\\
		Number of shells & 12 \\
		Diameters at principal plane & 293.76 -- 187.12 mm \\	
		Diameters at entrance pupil & 300 -- 191.1 mm\\
		Diameters at exit pupil & 274.88 -- 175.09 mm\\
		On-axis inc. angles at the intersection plane & 0.60 -- 0.39 deg\\
		Parabolic mirror lengths & 300 mm\\
		Hyperbolic mirror lengths & 300 mm \\
		Mirror wall thickness & 1.1 -- 0.65 mm\\
		Spider struts & 12\\
		Spider on-axis obstruction & 10\%\\
		Total weight of the mirrors & 41 kg\\
		\noalign{\smallskip}\hline
	\end{tabular}
\end{table}

\section{Experimental setup at PANTER}
\label{setup}
\begin{figure}
  	\includegraphics[width=0.75\textwidth]{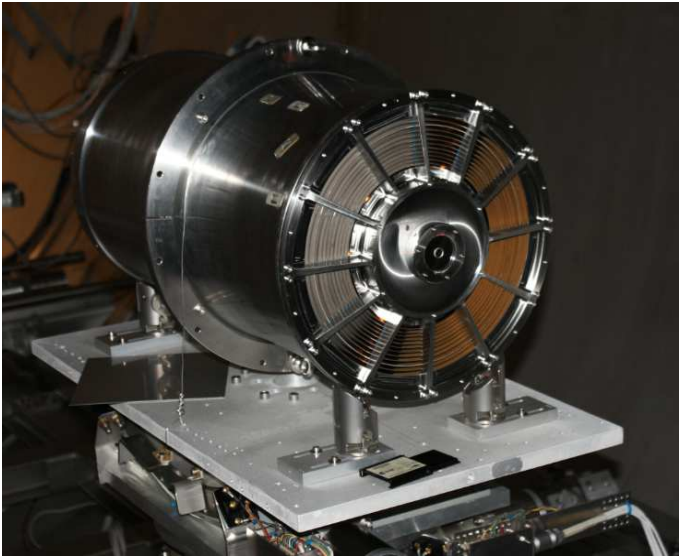}
	\caption{The JET-X FM2 mounted on the built-in manipulator at PANTER, seen from the aperture pupil. The module is laid upon 4 wheels to adjust the orientation of the spider struts.}
	\label{fig:FM2}  
\end{figure}

The JET-X flight modules include 12 Wolter-I mirror shells of decreasing diameter (Tab.~\ref{tab:JETXfacts}) and a 3.5 m focal length for a source at astronomical distance. The mirror wall thickness and the incidence angle for a source on-axis decrease proportionally to the diameter at the intersection plane (I.P.) between parabolic and hyperbolic segments, and the reflective coating is Gold. The mirror shells are integrated into a case with two ends made of 12 radial struts (Fig.~\ref{fig:FM2}) that support the shells, endowing them with the stiffness required to maintain the shape accuracy within 15 arcsec HEW. The fraction of effective area obscured by the spider struts is 10\%. In the experimental setup adopted at PANTER, the mirror module is simply mounted on the PANTER manipulator via an interface jig manufactured at INAF/OAB; the PANTER manipulator enables the lateral positioning to 1~$\mu$m accuracy and the rotation/tilt to 1~arcsec accuracy. This alignment accuracy is much better than requested to not affect the optical properties of the FM2.

The X-rays impinge on the module from the exit of a 123 m long, 1 m diameter vacuum tube. The X-ray sources are located in another vacuum chamber at the other end of the tube. The X-ray source produces unpolarized fluorescence lines, characteristic of the target material, plus a polarized Bremsstrahlung component. Most of the Bremsstrahlung emission is filtered out via suitable absorbers, but not entirely, therefore a small amount of polarization remains in the beam used to probe the optic. At the present time, the determination of the residual polarization is not relevant because the FM2 tests are performed with PANTER detectors that are insensitive to polarization. Moreover, at this stage, the tests with the GPD in the focus of the FM2 are solely aimed at measuring the angular resolution of the system. Several targets and filters are available to select monochromatic X-ray lines in the range of 0.18 -- 8.4~keV: in view of the utilization of the GPD \cite{Fabiani} to image the focal spot of the JET-X FM2, the interesting X-ray lines are the following: the C-K$\alpha$ (0.28 keV), the Al-K$\alpha$ (1.49 keV), the Ag-L$\alpha$ (2.98 keV), the Ti-K$\alpha$ (4.51 keV), the Fe-K$\alpha$ (6.40 keV), and the Cu-K$\alpha$ (8.05 keV).

Owing to the very long distance to the mirror module, the X-ray source at PANTER is representative of in-orbit conditions, but the finite distance results in a small but non-zero divergence that makes the optic to behave differently from the case of a source located at virtually infinite distance. The consequent effects are listed hereafter:
\begin{itemize}
\item{The incidence angles on the parabola and the hyperbola for a source on-axis, which should be identical for a source at infinity, now differ by twice the divergence angle.}
\item{A 15\% fraction of rays reflected by the parabolic segments near the maximum diameter misses the second reflection also with the source on-axis, reducing the number of focused rays.}
\item{The image is formed at a distance larger than the module focal length $f$. The image distance is $f' = f\,D_{\mathrm{sm}}/(D_{\mathrm{sm}}-f)$, where $D_{\mathrm{sm}}$ denotes the source-module distance.}
\item{Even at the displaced focal plane, the focusing suffers from a small aberration causing a modest increase of the HEW. This small effect is visible only with optics with a very high angular resolution, like Chandra.}
\end{itemize} 
A more detailed description of the finite distance effects is reported in \cite{Basso}. In the case of the JET-X FM2, the effects are quantified in Tab.~\ref{tab:findist}: they have been taken into account when comparing the measured optic performances to the theoretical predictions, especially the effective area values.

\begin{figure}
 	 \includegraphics[width=0.75\textwidth]{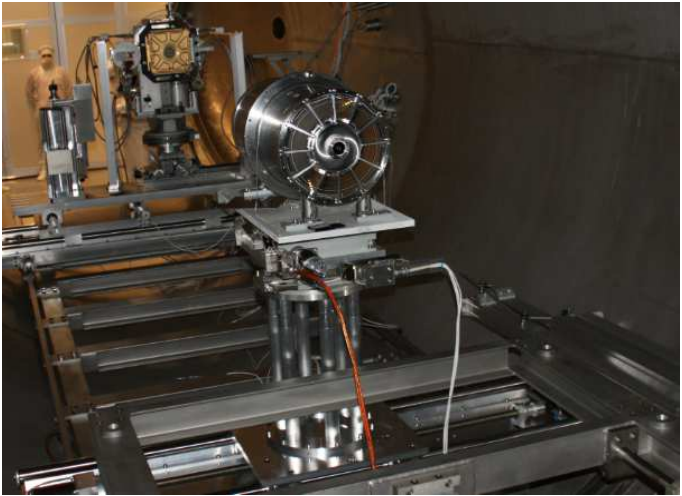}
	\caption{Experimental setup at PANTER. The PSPC and the TRoPIC detectors are visible in the background, near the exit of the vacuum tank.}
	\label{fig:setup2}      
\end{figure}

\begin{table}
\caption{Computed effects related to the finite distance of the X-ray source at PANTER. Even if 15\% of rays reflected by the parabolae misses the reflection on the hyperbola, the increased first reflection angle also increases the geometric cross-section of the parabolae by 10\%, partly compensating the effective area loss \cite{Spiga11}.}
\label{tab:findist}
	\begin{tabular}{ll}
		\hline\noalign{\smallskip}
		FM2 centre distance to X-ray source	& 128200 mm\\
		X-ray divergence at the I.P. with the source on-axis & 0.066 -- 0.042 deg\\
		Incidence angles on the parabolae at the I.P., source on-axis & 0.67 -- 0.43 deg\\
		Incidence angles on the hyperbolae at the I.P., source on-axis & 0.53 -- 0.35 deg\\
		On-axis parabola length lost fraction for double reflection & 15\%\\
		On-axis effective area lost fraction for double reflection & 5.5\%\\
		Intrinsic HEW blurring & 0.5 arcsec \\	
		Shifted focal distance & 3599 mm\\
		\noalign{\smallskip}\hline
	\end{tabular}
\end{table}

The focused rays have been recorded on the focal plane (Fig.~\ref{fig:setup2}) by two of the detectors available at PANTER \cite{Burwitz}:
\begin{itemize}
\item{The PSPC (Position Sensitive Proportional Counter), a gas detector developed for the ROSAT telescope. This detector has a circular sensitive area with a diameter of 8 cm, moderate spectral (40\% at 0.93 keV) and spatial (250 $\mu$m) resolutions, and an excellent linearity also at high count rates, without being affected by pile-up. This makes the PSPC suitable for the alignment of the optic and the effective area measurements.}
\item{TRoPIC, a solid state CCD camera with the same characteristics, but reduced size, of the imaging CCDs being produced for the eROSITA focal planes \cite{Meidinger}. TRoPIC is characterized by a smaller sensitive area (19.2 mm $\times$ 19.2 mm) than the PSPC, but higher spatial resolution (75 $\mu$m pixel size), possibly increased further to a sub-pixel resolution \cite{Dennerl}. However, at a 3600 mm distance from the optic, 1 pixel is equivalent to 4.3~arcsec, therefore a PSF with a 15 arcsec HEW is well oversampled also without sub-pixel analysis.}
\end{itemize}

\begin{figure}
 	 \includegraphics[width=0.75\textwidth]{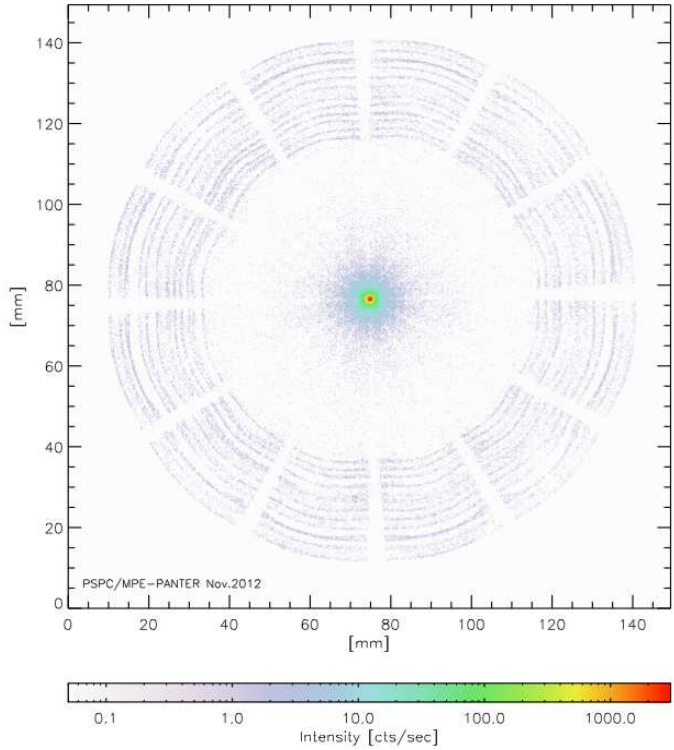}
	\caption{Mosaic of 9 exposures of the focal spot of the FM2, at 1.49 keV, taken with the PSPC. The double reflection spot is in the centre, the 12 concentric coronae are the  rays reflected once because of the finite distance of the X-ray source. Small roundness imperfections in the single reflection patterns stem from roundness errors of the mirror shells, however with a very small impact on the final HEW.}
	\label{fig:mosaic}      
\end{figure}

After the FM2 mounting and the tank evacuation, the mirror module has been aligned with the X-ray source on-axis. This has been done by centering the 12 single-reflection patterns (i.e., the rays that missed the second reflection) on the double reflection spot. Since the diameters of the coronae are larger than the PSPC diameter, the centering could not be checked with a single PSPC exposure. Therefore, 9 different, partly overlapping, exposures were taken and assembled to obtain a single picture (Fig.~\ref{fig:mosaic}) and so obtain a confirmation of the correct centering of the mirror module. The alignment was later refined using the Egger-Menz method described in \cite{Menz}. The best focal plane position was subsequently found by minimizing the HEW at the sole X-ray energy of 1.49 keV, with a $\pm$ 1~cm uncertainty. However, the distance of the best focal plane from the I.P. was not measured. The focal plane position, as recorded by the PANTER encoders, was used as a reference to locate the GPD in the subsequent phase, without moving the FM2.

The effective area measurements have been performed by firstly exposing the PSPC to the direct beam at one of the X-ray energies of interest, $E$. The average count rate of the direct beam, $C_{\mathrm d}(E)$, is thereby measured over the PSPC area, $A_{\mathrm d}$. Then the measurement of the focused intensity, $C_{\mathrm f}(E)$, is obtained with the PSPC in the focus of the FM2. In order to average the shadowing effects of the PSPC's entrance window mesh, the measurement is done in a slightly extra-focal position and the detector itself is scanned orthogonally to the focused beam. The measurement took nearly 1 hour to achieve a sufficient statistical accuracy: the effective area of the FM2 is thereby computed as 
\begin{equation}
	A_{\mathrm{M}}(E) = A_{\mathrm d}\frac{C_{\mathrm f}(E)}{C_{\mathrm d}(E)}\left(\frac{D_{\mathrm{sm}}}{D_{\mathrm{sd}}}\right)^2,
	\label{eq:EA}
\end{equation}
where $D_{\mathrm{sm}}$ is the source-mirror distance and $D_{\mathrm{sd}}$ is the source-detector distance. The last factor accounts from the divergence of the direct beam. Finally, since the direct and the focused beams have been recorded at different times and the source might have drifted in intensity over a long exposure, the intensity variation has been continuously controlled via a dedicated beam monitor. The count rates as determined for Eq.~\ref{eq:EA} have therefore been corrected in post-processing at PANTER (in the present case the correction amounted to less than 1\%). This standard procedure has been iterated for all the X-ray energies used during the test campaign. 

The HEW of the focal spot at the best focus has been measured with TRoPIC. In this case, the X-ray flux had to be kept at a very low count rate to have a negligible level of pile-up \cite{Ballet}. More exactly, the count rate recorded by TRoPIC was below 0.05~counts per frame (read every 50~msec), which results in a fraction of pile-up events below 0.08\% of the total, according to Poisson statistics and assuming a HEW of 15~arcsec. Hence, the integration time for every X-ray energy has been longer (nearly 100 minutes) than the one used for the effective area measurement, in order to achieve a statistical accuracy sufficient to reduce the HEW uncertainty below 1 arcsec. 

\section{Measurement results}
\label{results}
\subsection{Effective areas}
\label{results:EA}
The results of the effective area measurements are summarized in Tab.~\ref{tab:effarea}. The measurements have been performed with the optical module on-axis, 10 arcmin off-axis, and 15 arcmin off-axis (in the last case, only a couple of X-ray energies were used), with an uncertainty in the effective area values near 1\%. The measured values can be compared with the values originally measured on-axis \cite{Citterio96} at the energies of 1.49 and 8.05 keV (Tab.~\ref{tab:effarea_comp}). At 1.49 keV, the effective area is slightly smaller than the one measured in the 1996 campaign, but the difference is still of a few percent, on the order of the statistical error. The situation is different at 8.05 keV, where the effective area measured in this work is lower than the reported one in 1996, but the difference is now an order of magnitude larger than the statistical error. That difference, although significant, might be related to a modest surface roughness increase over the time elapsed since the first characterization.

\begin{table}
\caption{Effective area measurement results.}
\label{tab:effarea} 
	\begin{tabular}{|l|llllll|}
		\hline\noalign{\smallskip}
		X-ray energy & 0.28 keV & 1.49 keV & 2.98 keV & 4.51 keV & 6.40 keV & 8.05 keV \\
		\hline\noalign{\smallskip}
		On-axis effective area (cm$^2$)& 156.2 $\pm$ 1.9 & 147.1 $\pm$ 1.7 & 108.4 $\pm$ 1.2 & 112.1 $\pm$ 1.3 & 96.3 $\pm$ 1.2 & 52.6 $\pm$ 0.6 \\
		10 arcmin off-axis EA (cm$^2$) & 122.4 $\pm$ 1.6 & 119.1 $\pm$ 1.5 & 85.5 $\pm$ 1.0 & 86.8 $\pm$ 1.0 & 66.0 $\pm$ 0.8 & 27.6 $\pm$ 0.3 \\
		15 arcmin off-axis EA (cm$^2$) & 106.0 $\pm$ 1.4 & 101.6 $\pm$ 1.3 & -- & -- & -- & -- \\	
		\noalign{\smallskip}\hline
	\end{tabular}
\end{table}

\begin{table}
\caption{Comparison between the effective area values, as measured in 1996 and in the present campaign.}
\label{tab:effarea_comp} 
	\begin{tabular}{|l|ll|}
		\hline\noalign{\smallskip}
		X-ray energy &1.49 keV &  8.05 keV \\
		\hline\noalign{\smallskip}
		Measured effective area in 1996 (cm$^2$) & 152.0 $\pm$ 0.8 & 58.5 $\pm$ 0.5  \\
		Measured effective area in 2012 (cm$^2$) & 147.1 $\pm$ 1.7 & 52.6 $\pm$ 0.6  \\
		\noalign{\smallskip}\hline
	\end{tabular}
\end{table}

\begin{figure}
	\begin{tabular}{ll}
 	 \includegraphics[width=0.5\textwidth]{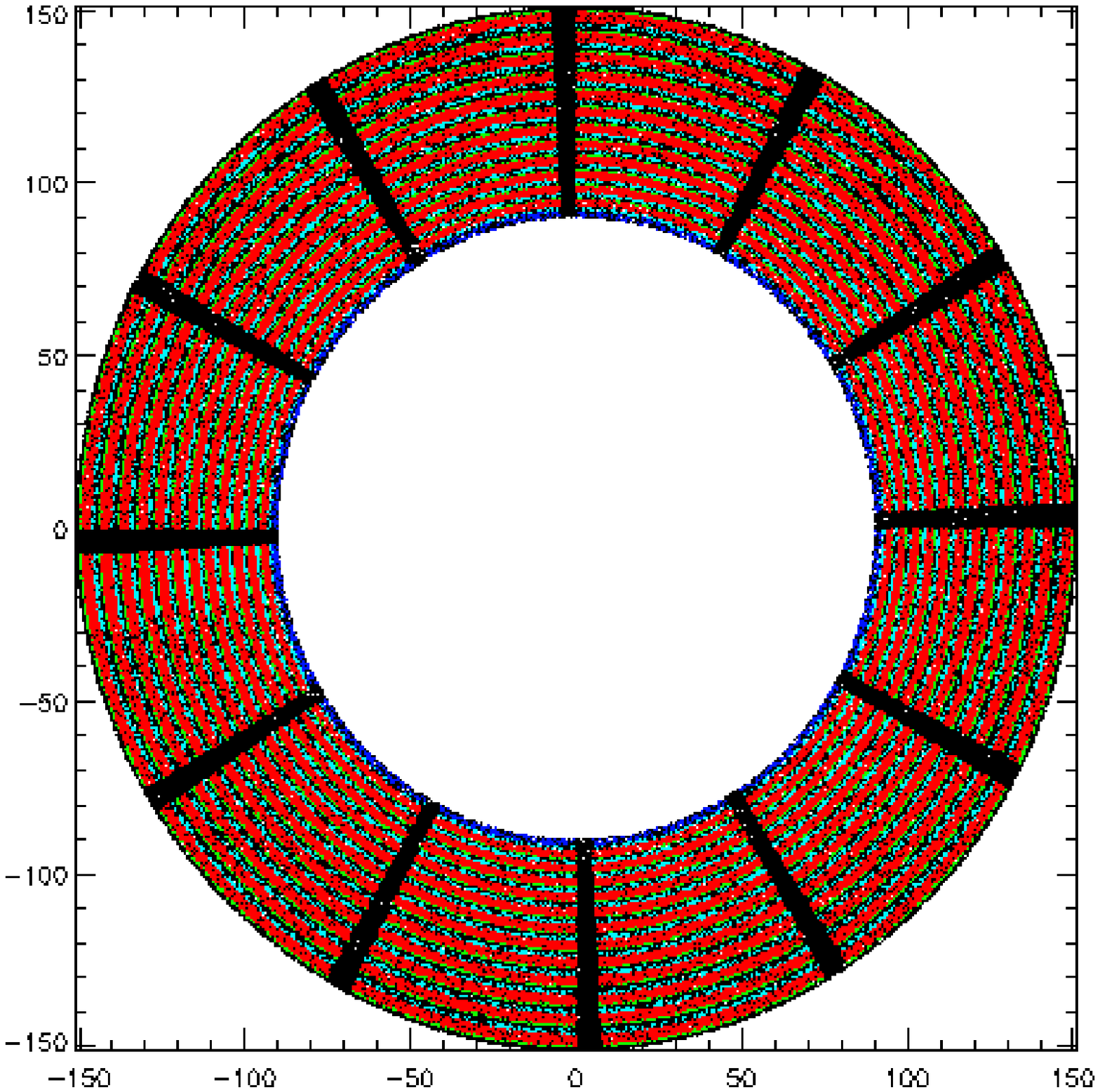} &  \includegraphics[width=0.5\textwidth]{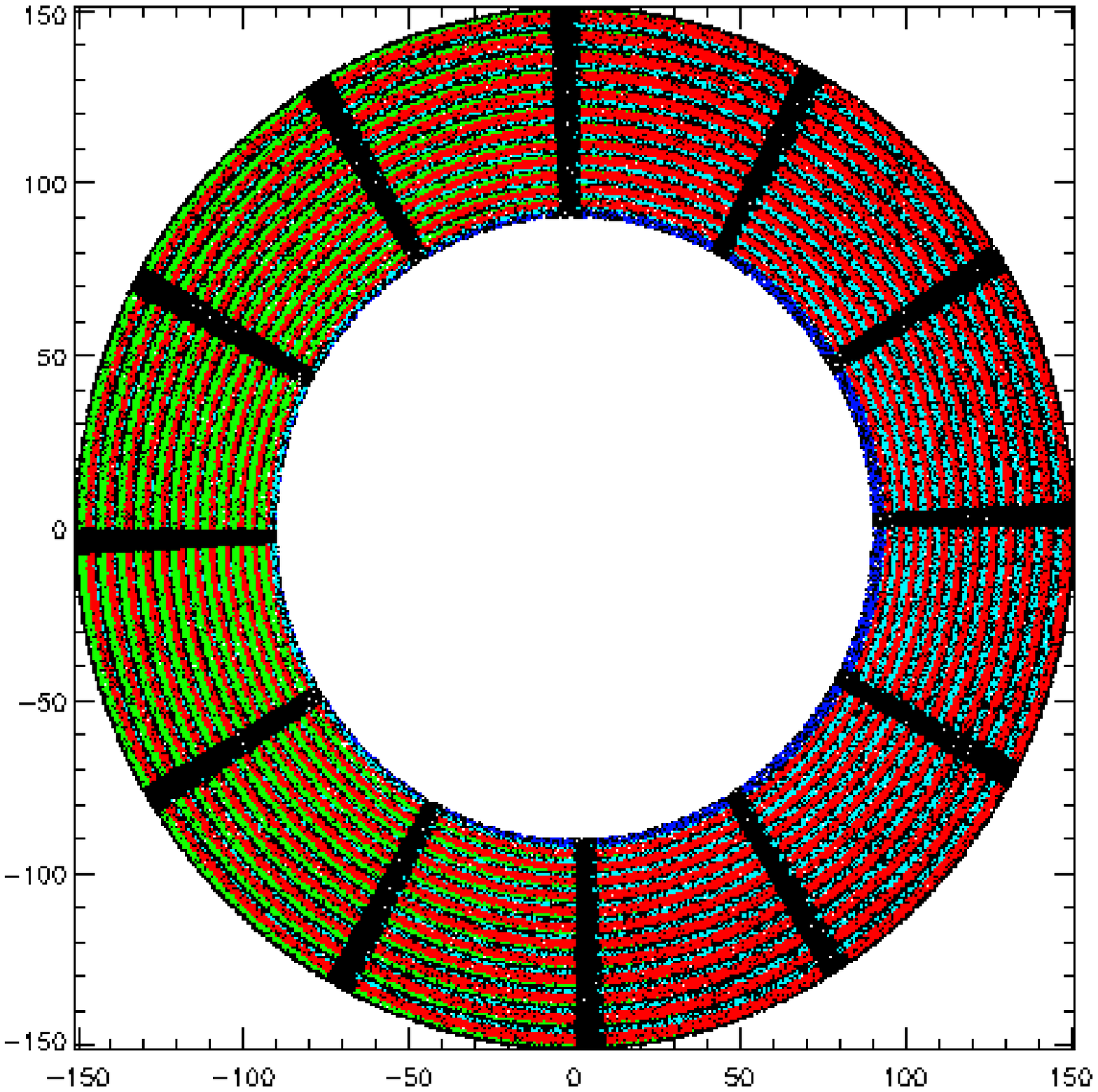}
	 \end{tabular}
	\caption{Ray-tracing simulation, with the X-ray source set (left) on-axis and (right) 10 arcmin off-axis. The ray locations are drawn of the entrance pupil of the FM2, with colors depending on their destiny. The doubly reflected rays are shown in red, the singly reflected rays are in green (parabola) or in cyan (hyperbola). The black zones are the rays obstructed or absorbed by the reflecting surfaces.}
	\label{fig:RT}      
\end{figure}

\begin{figure}
 	 \includegraphics[width=1.0\textwidth]{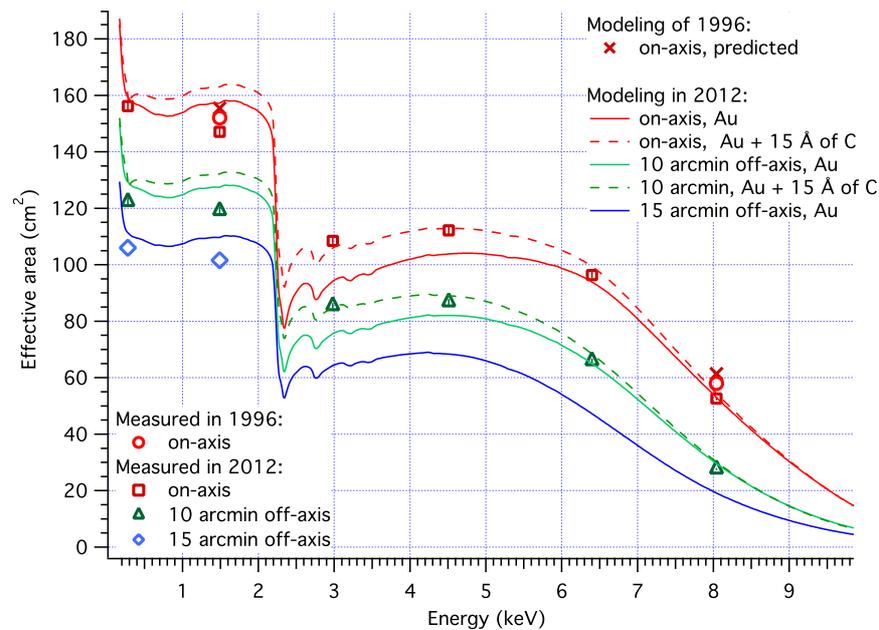}
	\caption{Measured effective area values of the FM2 (symbols). The measurements are compared to a computed coating model with bare Gold (solid lines) or Gold with Carbon overcoating (dashed line), using the X-ray optical constants of Gold by Palik~\cite{Palik}, of amorphous Carbon by Windt~\cite{Windt}, and assuming a roughness rms of 4.5~\AA. The density of the evaporated Gold was assumed to be 95\% of the nominal value.}
	\label{fig:Effarea}      
\end{figure}
As expected, the measured off-axis area is always lower than the measurement on-axis. This is related not only to the change of the incidence angles and the consequent change of reflectivity, but also to the increase of the number of rays that have missed the second reflection or had the first reflection on the hyperbolic segments, at the expense of the number of rays reflected twice and focused (Fig.~\ref{fig:RT}). Detailed computations \cite{Spiga11} have been performed to model the  effective area dependence on the X-ray energy. The modeling result, compared with the measured values, is plotted in Fig.~\ref{fig:Effarea}. Assuming a bare Gold coating (solid lines), the best match of data to model is found for a roughness rms of 4.5~\AA, Vs. a value of 3.5~\AA~inferred in \cite{Citterio96}. This value refers to the range of surface spatial wavelengths of a few tenths micron or shorter, i.e., the spatial frequencies causing X-ray scattering at sufficiently large angles to scatter X-rays out of the scanned region of the PSPC. However, even a zero-roughness model would underestimate the experimental results in the 2-5 keV energy range, just above the M-edge of Gold and therefore severely affected by photoelectric absorption. The higher reflectivity can be interpreted as a photoabsorption mitigation caused by a very thin hydrocarbon contamination layer -- of unknown origin and composition --  deposited onto the Gold layer. This effect was observed on the Chandra telescope optics \cite{Zhao} in orbit since 1999. Also a deliberate deposition of a low-Z material overcoating was suggested to enhance the mirror effective area in soft X-rays  \cite{Pareschi}. In the present case, the effect of hydrocarbon contamination can be simulated by a thin (15 \AA) amorphous Carbon layer on top of the Gold coating. The effective area computation (dashed lines in Fig.~\ref{fig:Effarea}) shows a much better matching to data between 2 and 5 keV. However, the measured area at 1.49~keV remains smaller than expected from the simulation: this might be related to the uncertain composition of the contamination layer.

\subsection{Point Spread Functions}
\label{results:PSF}
\begin{figure}
	\begin{tabular}{c}
 	 \includegraphics[width=0.8\textwidth]{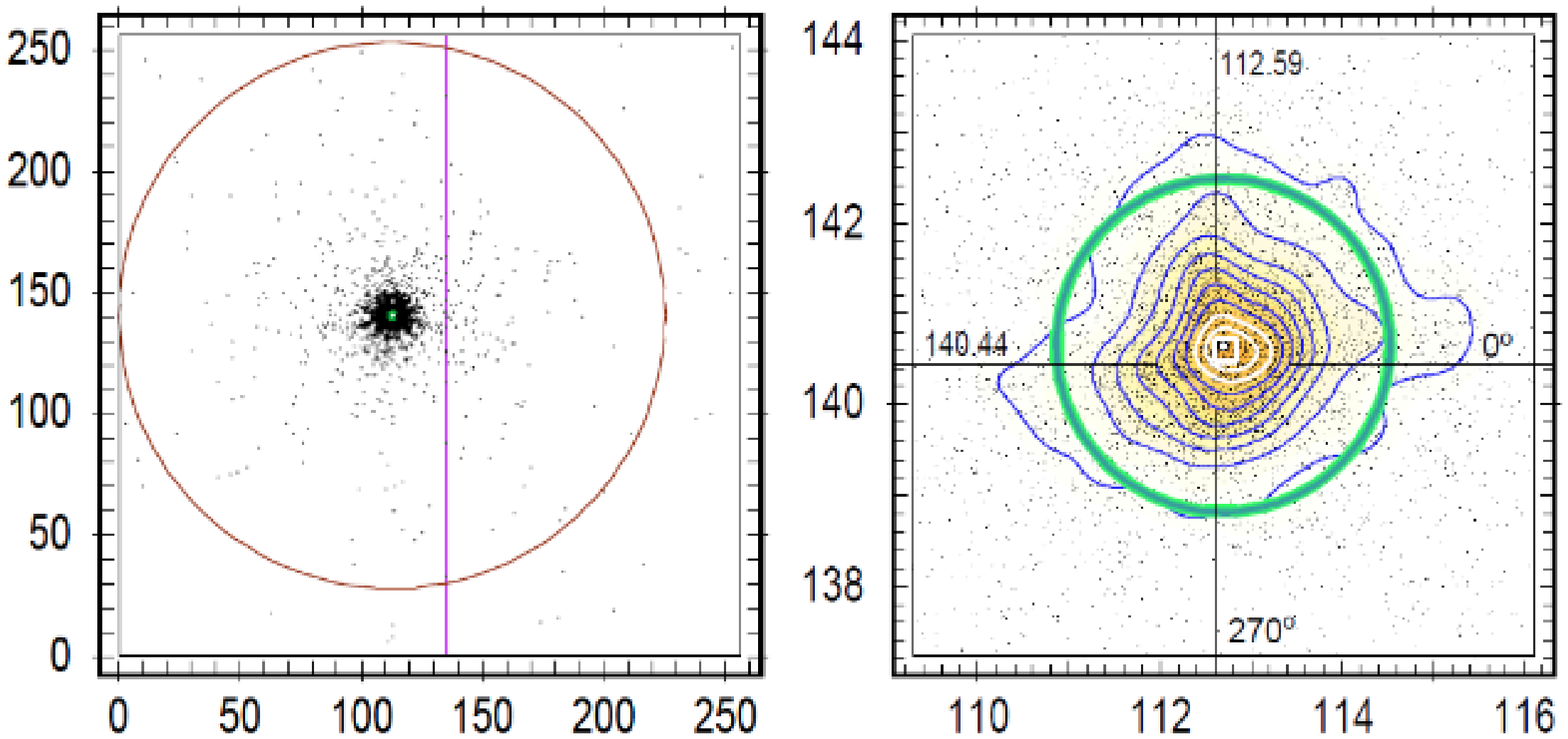}\\
	 \includegraphics[width=0.8\textwidth]{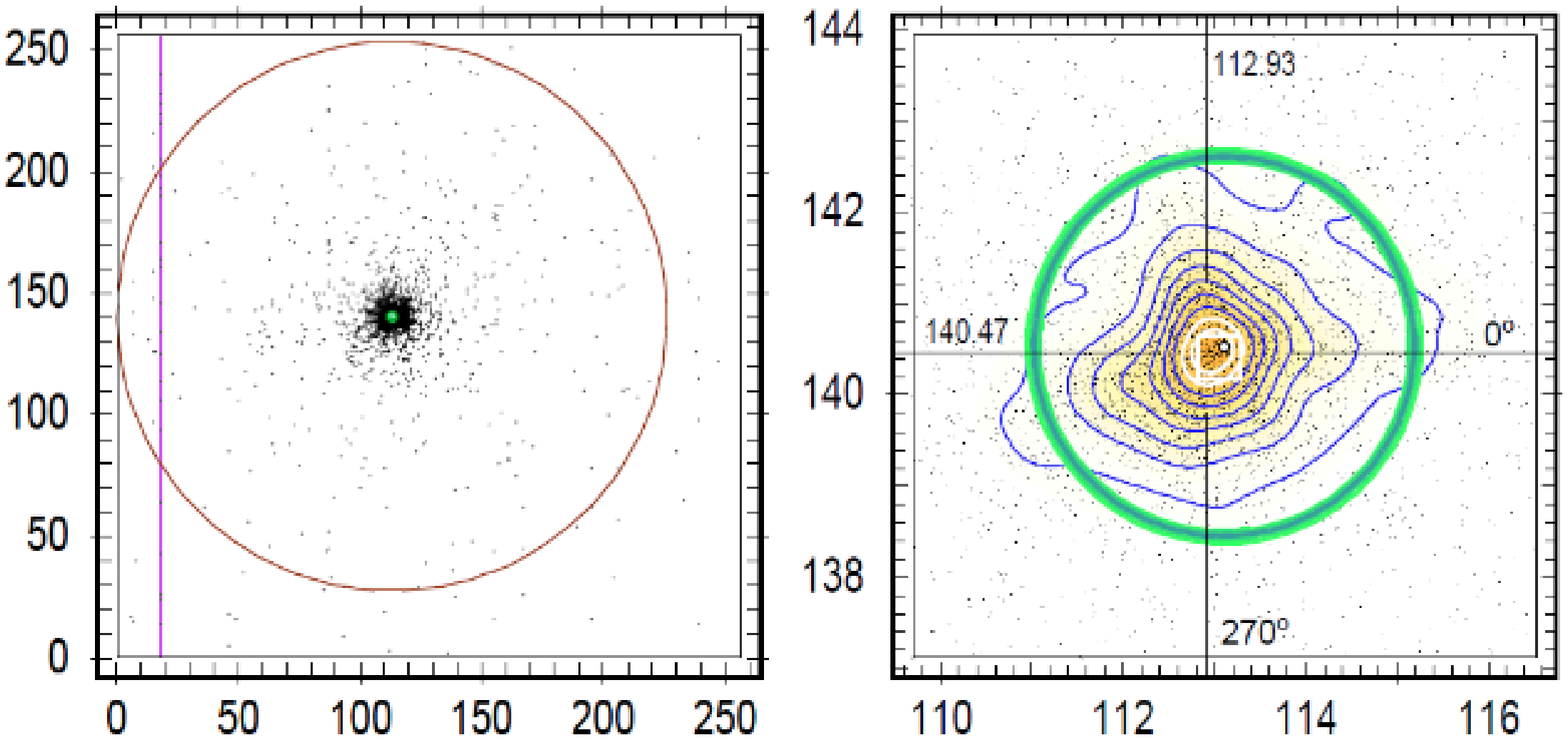}\\
	 \includegraphics[width=0.8\textwidth]{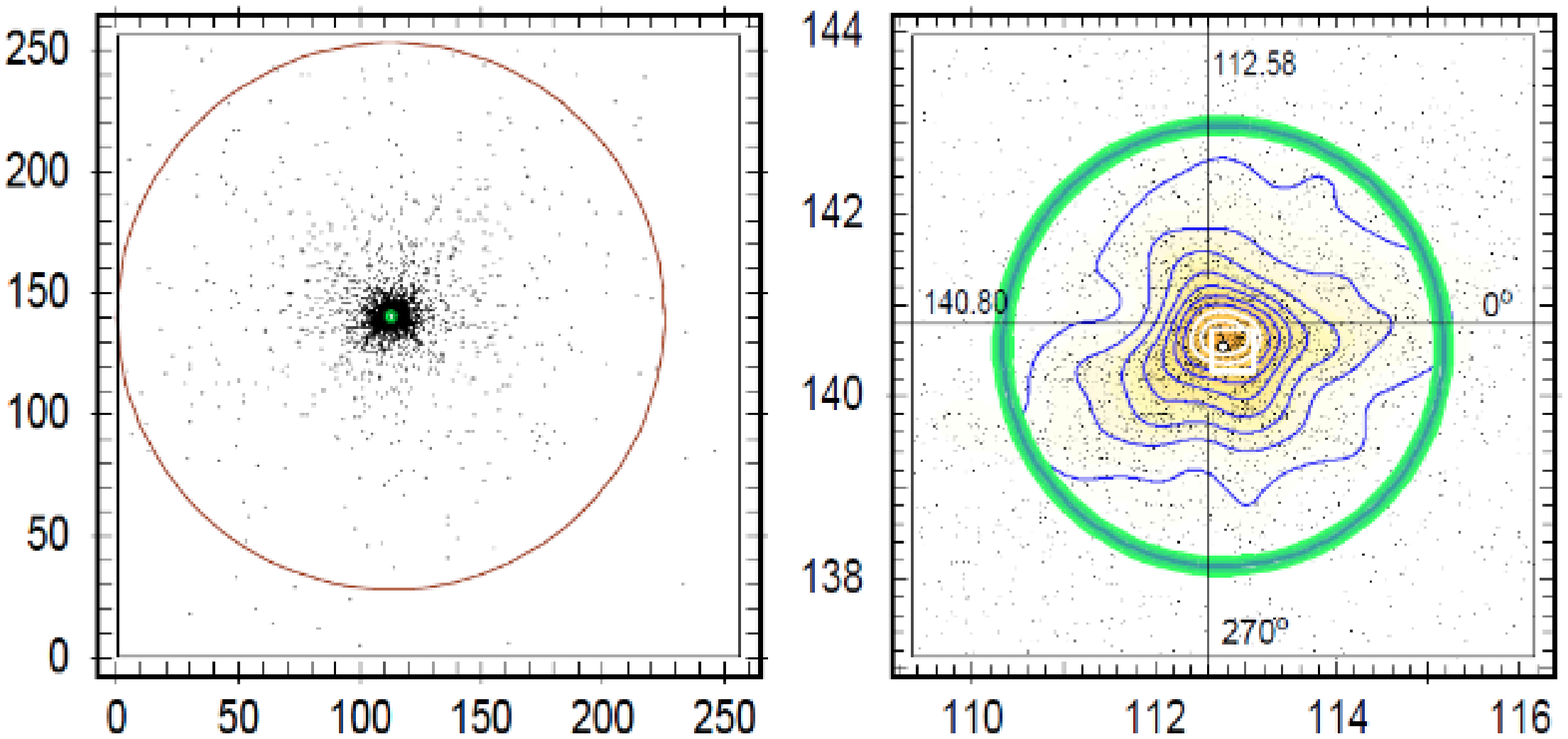}
	 \end{tabular}
	\caption{
Fig. 8 In focus PSF measurements of JET-X FM2 taken with the TRoPIC CCD camera at the X-ray energies 1.49 keV (top), 4.51 keV (middle), and 8.05 keV (bottom).
Left column: full detector sensitive area (256 $\times$ 256 pixel = 19.2 mm $\times$ 19.2 mm). The circle indicates the region over which the PSF has been computed. The columns in the upper two panels mark faulty columns of the detector that are not counted. Right column: Zoomed views of the PSF core, respectively. The HEW circles are traced in green.}
	\label{fig:PSFs}      
\end{figure}

The focal spots of the JET-X FM2 at 1.49 keV, 4.51 keV, and 8.05 keV (respectively the minimum, the average, and the maximum X-ray energy utilized to measure the PSF in focus) are displayed in Fig.~\ref{fig:PSFs}. In the left panels, the entire field of view of the TRoPIC detector is shown: the focal spots appear to be symmetric and entirely included in the TRoPIC detector sensitive area. The focal spot was slightly set off-center due to particular detector properties, hence the PSF -- obtained by integration of the image over circular coronae centered on the image centroid -- could be computed over a circular region of a diameter (16 mm) smaller than the TRoPIC side (19.2 mm). The PSF, once integrated over the radial coordinate, returned the HEW values reported in Tab.~\ref{tab:HEW_comp}, compared with the originally measured values in 1996 \cite{Citterio96}. A close view on the focal spots at the aforementioned energies is shown in the right panels of Fig.~\ref{fig:PSFs}. We point out that, owing to the appropriate long integration time, a very good statistical accuracy for the count rates has been achieved, which entails HEW uncertainties smaller than 1~arcsec.

A first glance at the HEW reported in Tab.~\ref{tab:HEW_comp} shows that, as expected, the HEW increases with the X-ray energy. This is a well known effect of the X-ray scattering stemming from the mirror surface roughness, which increases in relevance as the X-ray energy is increased. The X-ray scattering effect contributes to broaden the PSF, in addition to mirror figure errors. At very low X-ray energies, the X-ray scattering is negligible and the HEW is solely related to the mirror deformations, mostly longitudinal. Hence, the HEW value measured at 1.49 keV is basically figure error, while the gradual increase of HEW with the energy is caused by scattering. This is also visible from the contour plots in Fig.~\ref{fig:PSFs}, right: at the three energies shown, the focal spot is essentially unchanged in shape, but the HEW is degraded at higher energies because the scattering has increased the number of rays scattered out of the PSF core.

\begin{table}
\caption{Comparison between the angular resolution HEW, as measured in 1996 (with a MOS detector, $\sim$ 1 cm$^2$ area) and in the present campaign (TRoPIC, over a $\sim$ 2 cm$^2$ area).}
\label{tab:HEW_comp} 
	\begin{tabular}{|l|lllll|}
		\hline\noalign{\smallskip}
		X-ray energy & 1.49 keV & 2.98 keV & 4.51 keV & 6.40 keV & 8.05 keV \\
		\hline\noalign{\smallskip}
		HEW measured in 1996 (arcsec) & 14.6 & -- & -- & -- & 18.8 \\
		HEW measured in 2012 (arcsec) & 15.7 $\pm$ 0.4  & 16.8 $\pm$ 0.4 &18.0 $\pm$ 0.6 & 20.4 $\pm$ 0.9 & 20.8 $\pm$ 0.7 \\
		\noalign{\smallskip}\hline
	\end{tabular}
\end{table}

The HEW values are very close to the ones originally measured in \cite{Citterio96}, even if they are higher by 1 arcsec at 1.49 keV and by 2 arcsec at 8.05 keV (Tab.~\ref{tab:HEW_comp}). In both cases, the HEW increase is comparable to the measurement uncertainty, but still significant. The reason could be that the 1996 measurement was performed with a MOS detector (1 cm $\times$ 1 cm), while the present HEW values were computed over a circular region of an area of $\sim$ 2~cm$^2$:  the smaller detector cuts off the PSF wings, thereby returning lower HEW values. The difference is anyway negligible for the final scope of the experimental campaign, i.e., to ascertain the contribution to the GPD to the angular resolution of the polarimeter prototype. Since the GPD area is 2.25 cm$^2$ \cite{Fabiani}, the region of interest selected to compute the HEW values was correctly chosen to return HEW values that are directly comparable with the ones measured with the GPD in the focus of the JET-X FM2. 

\begin{figure}
 	 \includegraphics[width=1.0\textwidth]{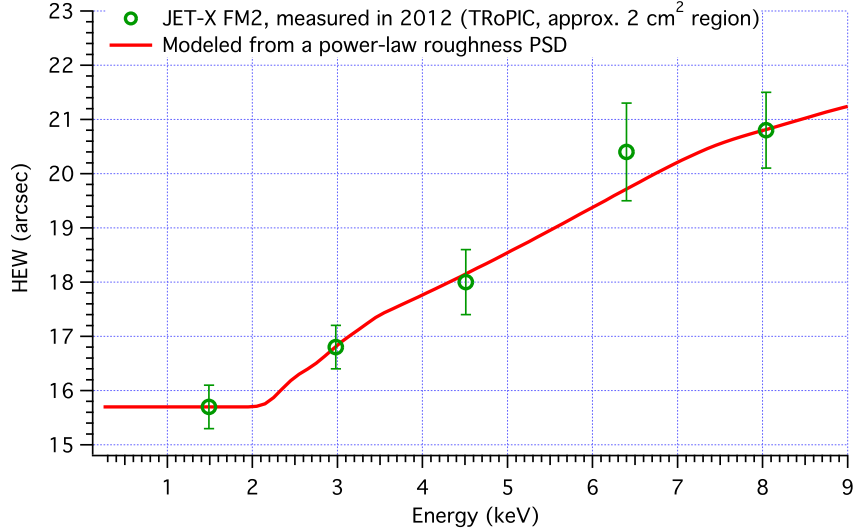}
	\caption{The measured HEW values (symbols), compared with the HEW trend expected from a figure HEW = 15.7~arcsec and a surface roughness PSD modeled as a power law with a spectral index of $n$ = 1.8 and a $K_n$ = 1.55~nm$^3 \mu$m$^{-1.8}$.}
	\label{fig:HEW_modeling}      
\end{figure}

A modeling of the HEW trend for increasing energy has been made by adopting a power-law roughness PSD (Power Spectral Density), as often done with superpolished surfaces \cite{Church}:
\begin{equation}
	P(f) = \frac{K_n}{f^n},
	\label{eq:powerlaw}
\end{equation}
where $f$ is the surface spatial frequency, $n$ is a spectral index taking on values $1 < n < 3$, and $K_n$ is a normalization constant. Starting from this assumption and taking $n$ and $K_n$ as free parameters, it is possible to reproduce the expected X-ray scattering HEW increase as a function of the X-ray energy and the incidence angle on the mirror shell \cite{Spiga}. This computation has been done in the 1-10 keV interval for all 12 mirror shells of the JET-X FM2, adding linearly to each of them the measured HEW value at 1.49 keV, assumed to be figure error \cite{RaiSpi}. The resulting HEW trends have subsequently been averaged properly \cite{Spiga2009} to return the expected values of the FM2 HEW. The result of the computation is displayed in Fig.~\ref{fig:HEW_modeling}, compared with the measured values. The best match between data and model is found for a spectral index $n$~=~1.8 and a normalization constant $K_n$~= 1.55~nm$^3 \mu$m$^{-1.8}$. 

Even if we have no measurements of the surface smoothness of the FM2 at hand, we notice that at these energies, the HEW degradation due to X-ray scattering is mostly determined by the spatial wavelengths in the millimeter range (i.e., causing small-angle scattering) more than in the micron range (i.e., causing scattering beyond the detector edge). In contrast, the high-frequency part of the spectrum is involved in the PSD normalization and therefore affects to some extent the HEW value used to compute the telescope absolute sensitivity, but this is not essential to the comparison with the angular resolution that has been measured with the GPD. Therefore, the power-law parameters we have derived are valid in the spectral window 1 mm -- 10 $\mu$m of spatial wavelengths, in which the inferred PSD returns a reasonable rms value of 7~\AA. The roughness rms derived from effective area measurements (4.5~\AA, Sect.~\ref{results:EA}) is essentially referred to a much higher frequency region of the PSD, also because the PSPC detector has a much larger area than TRoPIC. To return that rms at high frequencies, the PSD cannot exhibit a power-law up to high frequencies; rather, it is expected to be much flatter (i.e., to locally have a lower spectral index). As a matter of fact, a steep PSD at low frequencies ($n \approx 2$) followed by a flatter PSD at higher frequencies ($n \approx 1$) is a power spectrum often encountered in the surface roughness of replicated mirror shells \cite{Sironi}. 

\section{Conclusions}
\label{Conclusions}
Sixteen years after the initial calibrations, the JET-X FM2 has been re-tested at the PANTER X-ray facility. The main scope of the measurement campaign was to build a first prototype of imaging polarimetric telescope, coupling a high-performance optical module to a focal plane instrument with imaging and polarimetric capabilities at the same time (GPD). The tests were hitherto performed with a non-polarized source, aiming at measuring the angular resolution of the optics + GPD system. In this paper we have reported the results of the tests performed on the sole JET-X optical module using the imaging detectors available at PANTER. We have verified that the effective area values and the angular resolution have changed only slightly over 16 years of storage. Even if the 1996 data do not allow a comparison in the 2-5 keV energy range, we have found an effective area excess in this band that we interpret as a slight contamination of hydrocarbon on the Gold surface. The origin of this contaminant is unknown, because the FM2 was always kept stored in pure Nitrogen atmosphere. We expect this layer to have grown extremely slowly over time, therefore no effective area change should have occurred between the FM2 test and the subsequent characterization of the GPD in its focus. However, this result should be considered in future calibrations of X-ray optical modules, especially if a considerable time has elapsed since the manufacturing. We have also provided a modeling of the surface roughness in the millimeter range of spatial frequencies, in order to explain the HEW degradation with increasing energy, in terms of the surface roughness power spectrum. The results of the second part of the campaign, with the GPD in the focal plane of the JET-X FM2, is reported in another paper \cite{Fabiani}.

\begin{acknowledgements}
The JET-X Flight Modules development and this study are financed by ASI (Agenzia Spaziale Italiana).
\end{acknowledgements}

\end{document}